\documentclass[fleqn,usenatbib,twocolumn]{mnras}

\usepackage{newtxtext,newtxmath}

\usepackage[T1]{fontenc}
\usepackage{ae,aecompl}

\usepackage{graphicx}	
\usepackage{amsmath}	

\usepackage{amssymb}	
\usepackage{bm}
\usepackage{textcomp}
\usepackage{color}

\usepackage{ulem,times}

\newcommand{\mpch}{\,\mathrm{Mpc}/h} 
\newcommand{\hmpc}{\,h\mathrm{Mpc}^{-1}} 
\newcommand{\mpc}{\,\mathrm{Mpc}} 
\newcommand{\ic}{\mathrm{i}}
\newcommand{\fc}{\mathrm{f}}
\newcommand{\rec}{\mathrm{r}}
\newcommand{\Crit}{critical boundary}

\newcommand{\eg}{\textit{e}.\textit{g}., }

\title[BAO reconstruction]{Baryon acoustic oscillations reconstruction using convolutional neural networks}

\author[Tian-Xiang Mao et al]{
\parbox[t]{\textwidth}{
Tian-Xiang Mao$^{1,2}\thanks{E-mail: maotianxiang@nao.cas.cn}$, Jie Wang$^{1,2}\thanks{E-mail: jie.wang@nao.cas.cn}$, Baojiu Li$^{3}$, Yan-Chuan Cai$^{4}$, Bridget Falck$^{5}$, \\ Mark Neyrinck$^{6}$ and Alex Szalay$^{5}$}\\ \\
\parbox[t]{\textwidth}{
$^{1}$National Astronomical Observatories, Chinese Academy of Sciences, Beijing, 100012, China\\
$^{2}$University of Chinese Academy of Sciences, Beijing 100049, China\\
$^{3}$Institute for Computational Cosmology, Department of Physics, Durham University,
Durham DH1 3LE, UK\\
$^{4}$Institute for Astronomy, University of Edinburgh, Royal Observatory, Blackford Hill, Edinburgh EH9 3HJ, UK\\
$^{5}$Department of Physics and Astronomy, The Johns Hopkins University, Baltimore, MD 21218, USA\\
$^{6}$Ikerbasque, the Basque Foundation for Science and Dept.\ of Physics, University of the Basque Country, Bilbao, Spain
}}

\date{Accepted 2020 November 25. Received 2020 October 31; in original form 2020 February 24}

\pubyear{2020}

\begin{document}
\label{firstpage}
\pagerange{\pageref{firstpage}--\pageref{lastpage}}
\maketitle

\begin{abstract}
We propose a new scheme to reconstruct the baryon acoustic oscillations (BAO)
signal, which contains key cosmological information, based on deep
convolutional neural networks (CNN). Trained with almost no fine tuning, the
network can recover large-scale modes accurately in the test set: the
correlation coefficient between the true and reconstructed initial conditions
reaches $90\%$ at $k\leq 0.2 \hmpc$, which can lead to significant improvements
of the BAO signal-to-noise ratio down to $k\simeq0.4\hmpc$. Since this new
scheme is based on the configuration-space density field in sub-boxes, it is
local and less affected by survey boundaries than the standard reconstruction
method, as our tests confirm. We find that the network trained in one cosmology
is able to reconstruct BAO peaks in the others, i.e. recovering information
lost to non-linearity independent of cosmology. The accuracy of recovered BAO
peak positions is far less than that caused by the difference in the cosmology
models for training and testing, suggesting that different models can be
distinguished efficiently in our scheme. It is very promising that Our scheme
provides a different new way to extract the cosmological information from the
ongoing and future large galaxy surveys. 
\end{abstract}




\section{Introduction}

Understanding observations and using them to constrain the nature of physics is
a long-term task in modern cosmology, which requires both obtaining
high-quality data and developing accurate data analysis methods. The baryon
acoustic oscillations (BAOs), imprinted on large-scale structure, is a standard
ruler in cosmology and play an important rule in studying the cosmic expansion
history, and the properties of dark energy.

BAOs arise from the coupling of baryons and photons in the early Universe
\citep[for a recent review see][]{Weinberg2013}. After recombination, this
feature is imprinted in both the cosmic microwave background (CMB) and the
matter distribution. CMB anisotropy measurements have provided highly precise
constraints on the sound horizon at high redshift
(\cite{Peebles1970,Sunyaev1970}, and for more recent results see
\cite{Planck2016,WMAP9}). In the case of galaxy clustering, the BAO feature
imprinted in the form of a characteristic scale provides an absolute distance
scale, and can be used to map the expansion history $H(z)$. It has been
detected at redshift about $0.1\sim 0.8$ and $2.5$ \citep[for recent results
and a summary of BAO measurements, see {\it e.g.},][]{Alam2017} as a peak in
the correlation function or as a harmonic sequence of oscillations in the power
spectrum. Fortunately, the relatively large scale of the BAO feature (about
$145\mpc$) protects it from substantial nonlinearity, making it
straightforward to interpret, and a powerful tool for measuring the
cosmological distance scale.

However, late-time nonlinear evolution does broaden and shift the BAO peak in
the correlation function, or, equivalently, dampen high-$k$ oscillations in the
power spectrum, which decreases the accuracy and precision of the detection of
the BAO signal
\citep{Meiksin1999,Springel2005,Angulo2005,Seo2005,Jeong2006,Huff2007,Eisenstein2007a,Angulo2008,Padmanabhan2009,Seo2010,Mehta2011,Sherwin2012}.
Furthermore, a number of other effects can also introduce difficulties in BAO
measurement, such as the survey boundary, galaxy bias and redshift-space
distortions. In order to correct this blurring caused by nonlinear evolution,
\cite{Eisenstein2007b} proposed a reconstruction method (hereafter `standard
reconstruction') by moving the galaxies back along large-scale bulk flows,
which considerably enhances the BAO peak both in theory
\citep{Seo2008,Padmanabhan2009b,Noh2009,Seo2010,Mehta2011,White2015,Schmittfull2015}
and in observations
\citep{Padmanabhan2012,Xu2013,Anderson2014,kazin2014,Ross2015,Beutler2016,Beutler2017,Hinton2017}.

Recently, motivated by the success of standard reconstruction and the current
or upcoming observations (\eg 2MASS \citep{2MASS}, 4MOST \citep{4MOST}, 6dF
\citep{6dF}, SDSS \citep{Alam2017}, DES \citep{DES}, PFS \citep{PFS}, DESI
\citep{DESI}, EUCLID \citep{Euclid}, LSST \citep{LSST}, Tianlai
\citep{Tianlai}, CHIME \citep{CHIME}, HIRAX \citep{HIRAX}, BINGO \citep{BINGO},
and SKA \citep{SKA}), various other reconstruction methods have been proposed
and gained broader applications (for a review, see \cite{Schmittfull2017}). For
example, \cite{Zhu2016,Zhu2017} proposed a nonlinear reconstruction technique
based on iteratively solving the coordinate transform between the Lagrangian
and Eulerian frames. It has been tested for dark matter density fields
\citep{Zhu2016,Zhu2017}, and it has been investigated with respect to Fisher
information \citep{Pan2017}, BAO \citep{Wang2017}, biased tracers
\citep{Yu2017,Wang2018} and redshift-space distortions \citep{Zhu2018}.
\cite{Schmittfull2017} described an iterative method to reconstruct the initial
conditions, and \cite{Seljak2017,Feng2018,Modi2018} converted the
reconstruction to an optimization problem by forward modeling. For similar
purposes, \cite{Shi2018} proposed a multi-grid relaxation algorithm and
extended it for biased tracers \citep{Birkin2018} and to remove redshift-space
distortions from galaxy clustering \citep{Wang2019}. Most of these methods
achieve substantial improvements beyond the standard reconstruction and some
other methods have been designed to gain more information in some specific
cases
\citep[\eg][]{Kitaura2013,Wang2013,Jasche2013,Burden2015,Obuljen2017,Hada2018,Hada2018b,Sarpa2018,Bos2019,Kitaura2019,Leclercq2019}. 
 
Another potential approach to extract the BAO features from galaxy surveys is
by using neural networks \citep[hereafter networks; for some reviews
see][]{Lecun2015,Goodfellow2016}, which have been widely used in various fields
in astronomy, such as gravitational lensing
\citep{Springer2018,Tewes2018,Gupta2018,Li2018_NN,Morningstar2018,Morningstar2019},
the Cosmic Microwave Background \citep{Caldeira2018}, neutral hydrogen
\citep{Gillet2018,Shimabukuro2017,Rafieferantsoa2018,Pablo2020}, constraining
cosmological parameters
\citep{Mathuriya2018,Gupta2018,Ravanbakhsh2017,Schmelzle2017}, large-scale
structure classification \citep{Aragon-Calvo2018}, and generation
\citep{Rodriguez2018} and structure formation
\citep{Berger2018,Lucie-Smith2018,Lucie-Smith2019,Modi2018,He2019}.

In this paper, we propose a new, network-based method to reconstruct the BAO
signal from a dark matter density field. We use high-resolution N-body
simulations, which provide all necessary information to construct our network
model. In this case, we convert the reconstruction problem to a nonlinear
mapping from final nonlinear density to initial linear density, by introducing
a large number of parameters, which are optimized by feeding simulation data to
the network for training.

This is an independent method from the ones commonly used in BAO analyses, and
can be used to identify and understand potential modeling systematics in BAO
measurements. Unlike methods based on perturbation theory, our method is more
robust in regions close to the survey boundary, because it reconstructs the
initial linear density from the local final nonlinear density in configuration
space. In contrast, in Fourier space, the effect of the survey boundary is a
broad window function that is convolved onto the density field, which can have
a global impact on reconstruction. This effect becomes local in configuration
space, reducing the impact on our reconstruction.

This paper is organized as follows. In Section \ref{sec:method} we describe our
network model and the simulations used in this work. It is followed by results
of the reconstruction in Section \ref{sec:result}. We then discuss cosmology
dependence and survey boundary effects in Section \ref{sec:discussion}, and
finally conclude in Section \ref{sec:con}.

\section{Method}
\label{sec:method}

In this section, we first describe the reconstruction problem by way of maximum
likelihood estimation and show that the reconstruction can be represented by
the network. After that, we review traditional and convolutional neural
networks and describe the network model we used in this work. Finally, we show
the training process and describe the dataset used in this paper.

\subsection{Maximum likelihood estimation}
\label{sec:model}

In general, the nonlinear power spectrum of the final density field can be
expressed as a sum of two parts \citep{CrocceScoccimarro2006}: 
\begin{equation}
P_{\mathrm{nl}}(k)=G^2(k) P_\mathrm{ini}(k) + P_\mathrm{mc}(k),
\end{equation}
where the $P_\mathrm{ini}(k)$ is the linear power spectrum, $G(k)$ is the
propagator and $P_\mathrm{mc}(k)$ indicates the power spectrum from mode
coupling.  The propagator term encodes the memory of the initial conditions,
and $G(k)\rightarrow1$ as $k\rightarrow0$, indicating that the information on
large scales is well preserved even at late times. On intermediate scales, mode
coupling modulates large-scale information into smaller, nonlinear, scales. In
regions where shells cross, small-scale information is lost, and it cannot be
estimated uniquely if only the final density field is provided. Since the BAO
features are located on large and intermediate scales, shell crossing is not a
major concern for BAO reconstruction. As we focus on BAO reconstruction in this
paper, we shall, following the standard practice in the reconstruction
literature, assume there is no shell crossing.

We suppose a parameteric model $f(\delta_{\fc}; \boldsymbol{\theta})$, with
parameter set $\boldsymbol{\theta}$, that can predict the initial linear
density field $\delta_{\ic}$ above a certain length scale, given the
corresponding final density field $\delta_{\fc}$. It can be written as 
\begin{equation} \label{eq:datapoint}
\delta_{\ic} = f(\delta_{\fc}; \boldsymbol{\theta}),
\end{equation} 
and the corresponding conditional probability is
$P(\delta_{\ic}|\delta_{\fc};\boldsymbol{\theta})$. Under the assumption of
independent and identically distributed (i.i.d.) data points, the likelihood is
\begin{equation}
    \mathcal L = \prod\limits_{k=1}^{N_{\rm p}} P(\delta_{\ic}|\delta_{\fc};\boldsymbol{\theta}),
\end{equation}
where $N_{\rm p}$ is total number of data points, i.e., pixels or cells where
the density fields are evaluated, and the maximum likelihood estimation of
$\boldsymbol{\theta}$ is
\begin{equation}
    \hat{\theta}_{\mathrm{ML}}= \mathop{\arg\max}_{\boldsymbol{\theta}}
    \mathcal L= \mathop{\arg\max}_{\boldsymbol{\theta}} \sum_{k} \log
    P(\delta_{\ic}|\delta_{\fc};\boldsymbol{\theta}).
\end{equation}
If we assume
\begin{equation}
\label{eq:gauss}
P(\delta_{\ic}|\delta_{\fc};\boldsymbol{\theta})=\mathcal{N}(\delta_{\ic};f(\delta_{\fc};
\boldsymbol{\theta}),\sigma),
\end{equation}
where the measured (or simulated) $\delta_{\ic}$ is regarded as a data point
sampled from a Gaussian distribution $\mathcal{N}$, the function
$f(\delta_{\fc}; \boldsymbol{\theta})$ gives the prediction of the mean of
$\mathcal{N}$ and a fixed standard deviation $\sigma$ is assumed to simplify
the problem. In such conditions, the maximum likelihood estimation is
equivalent to minimizing the mean square error (MSE),
\begin{equation}
    \hat{\boldsymbol{\theta}}_{\mathrm{ML}}= 
    \mathop{\arg\min}_{\boldsymbol{\theta}} 
    \sum^k \left(f(\delta^k_{\fc}; \boldsymbol{\theta})
    -\delta^{k}_{\ic}\right)^2,
\label{eq:ML}
\end{equation}
where the superscript $k$ again indicates i.i.d. data points.  We assume a
Gaussian distribution of $\mathcal{N}$ in Eq.~(\ref{eq:gauss}) because there is
no preferred distribution. The Gaussian distribution is a natural choice
according to the central limit theorem, which applies when the number of data
points is large.

In the reconstruction problem, it is not easy to estimate
$\boldsymbol{\theta}_{\mathrm{ML}}$ directly because we do not know the
mathematical form of the model $f(\delta_{\fc}; \boldsymbol{\theta})$. A
possibility is fitting the parameters $\boldsymbol{\theta}$ from observed data
with the help of an optimization method. However, due to the complexity of the
model and the huge number of parameters, traditional optimization methods, such
as Markov chain Monte Carlo (MCMC) are not efficient. In this work, we tackle
this difficulty by using gradient-based neural network to construct model and
optimize parameters.

\subsection{Artificial neural networks}
\label{sec:ANN}

Artificial neural networks (see some reviews, \cite{Lecun2015, Goodfellow2016})
are suitable for solving problems with no known specific mathematical
expressions. By constructing a nonlinear parametric model, the network converts
complex problems into non-convex optimization and optimizes the trainable
parameters by gradient-descent based methods \citep[\eg stochastic gradient
descent,][]{SGD}. The process of optimizing trainable parameters by feeding a
series of data points into a fixed network architecture is called
\textit{training}. 

A standard feed-forward neural network consists of multiple layers. Each layer
performs a weighted linear combination of its inputs, followed by an
element-wise nonlinear activation function and a bias term. These weights and
biases on all layers constitute the trainable parameters of the network.  

For layer $n$, if we set the input vector as $\boldsymbol{x}_{n-1}$, weight
matrix $\mathbfss{W}_{n}$ and bias vector $\boldsymbol{b}_n$, then the output
of this layer is 
\begin{equation}
\label{eq:layer}
\boldsymbol{x}_n = a(\mathbfss{W}_n\boldsymbol{x}_{n-1}+\boldsymbol{b}_n).
\end{equation}
Here the $a$ denotes a nonlinear activation function. In this paper, we use the
rectified linear unit (ReLU, \cite{ReLU}) activation function. For the network,
the output of one layer is the input of the next layer. By stacking a series of
functions in Eq.~\eqref{eq:layer}, the network will have the potential to
approximate the $f(\delta_{\fc};\boldsymbol{\theta})$ in Eq.\eqref{eq:ML} and
the trainable parameters correspond to the parameter set $\boldsymbol{\theta}$.

In deep learning, increasing the number of layers $N$ always expands the
capacity of the network, by enlarging the hypothesis space of solutions that
the algorithm is able to choose from, although it may lead to difficulties in
training. Once the network architecture is determined, the trainable parameters
in the network will be optimized by minimizing a loss function. The loss
function describes a kind of distance between the network prediction and the
target value. In this paper, we choose MSE loss as the loss function.

\subsection{Convolutional neural networks}
\label{sec:CNN}

\begin{table*}
\caption{\label{tab:network} The network architecture. Our network consists
of 7 convolutional layers and one fully connected layer. Here, the kernel
size shows the shape of convolutional kernels in each convolutional layer. The
output shape describes the output size of each layer. For convolutional layers,
each dimension means [batch size, depth, height, width, channels]. For the
fully connected layer, each dimension indicates [batch size, channels]. In the
layer before fully connected, we average the output of the conv7 layer in
dimensions of performing convolution, which can also be seen as an average
pooling layer. All convolutional layers are followed by a ReLU \citep{ReLU}
activation function in our network.} 
\centering
\begin{tabular}{l|c|c|c|c}
\hline 
Layer & Kernel size & Output shape & Stride & Activation function \\
\hline \hline
input & None & (None, 39, 39, 39, 1) & None & None \\\hline
conv1 & (3, 3, 3) &(None, 20, 20, 20, 32)& (2, 2, 2)  & ReLU \\ \hline
conv2 & (3, 3, 3) &(None, 20, 20, 20, 32) & (1, 1, 1)  & ReLU  \\\hline
conv3 & (3, 3, 3) &(None, 10, 10, 10, 64) & (2, 2, 2)  & ReLU  \\\hline
conv4 & (3, 3, 3) &(None, 10, 10, 10, 64) & (1, 1, 1)  & ReLU \\\hline
conv5 & (3, 3, 3) &(None, 5, 5, 5, 128) & (2, 2, 2) & ReLU \\\hline
conv6 & (3, 3, 3) &(None, 5, 5, 5, 128)  & (1, 1, 1)  & ReLU \\\hline
conv7 & (1, 1, 1) &(None, 5, 5, 5, 128)  & (1, 1, 1)  &  ReLU  \\\hline
mean  & None &(None, 128)   & None  & None  \\\hline
fully connected & None &(None, 1) & None  & None  \\\hline\hline
\end{tabular}
\end{table*}

Convolutional neural networks \citep[hereafter CNNs, see \eg][]{CNN,CNN2} are
well known in processing visual imagery because of their shift-invariant
property and reduced number of free parameters. In this study, we will perform
the MSE estimation described in Eq.~\eqref{eq:ML} using a 3-D CNN. 

CNNs replace the matrix-vector product $\mathbfss{W}_n\boldsymbol{x}_{n-1}$ in
Eq.~\eqref{eq:layer} by a sum of convolutions, the latter being more efficient
and having fewer trainable parameters. Like in Eq.~\eqref{eq:layer}, we
represent the output of the $l$-th kernel in layer $n$ as
\begin{equation} 
    \boldsymbol{x}_n^{l} = a\left(\sum^{k}\mathbfss{W}_n^{l}\otimes 
\boldsymbol{x}_{n-1}^{k}+\boldsymbol{b}_n^{l}\right),
\end{equation} 
where $\otimes$ indicates the 3-D convolution operation, $\mathbfss{W}_n^{l}$
denotes trainable convolutional kernels for layer $n$, $l$ indicates the $l$-th
kernel in this layer and $k$ indicates the output corresponding to the $k$-th
convolutional kernel in the previous layer, which is also called the $k$-th
channel.

In addition to the convolutional layers, standard CNNs usually contain
\textit{pooling layers} \citep[\eg][]{CNN2}. In our network, the pooling layers
are replaced by a striding in $2$ voxels per side in the convolutional
calculation. For all convolutional kernels in this paper, we set their size as
$3\times3\times3$ voxels, except the last convolution layer whose kernel size
is $1\times1\times1$. The detailed network architecture is shown in Table
\ref{tab:network}. 

As described in Eq.~\eqref{eq:datapoint}, the input and output of the network
are the final density field $\delta_{\fc}$ and the initial density field
$\delta_{\ic}$, respectively. To further reduce the computational and memory
requirements in the training, we generate $\delta_{\fc}$ and ${\delta_{\ic}}$
fields in a small sub-box instead of the whole simulation box. $\delta_{\fc}$
is generated in a cubic region with a length of $76\mpch$ per side. For
$\delta_{\ic}$, we choose the corresponding central region at initial time with
a length of $1.95\mpch$ per side. This is because as the structure evolves, the
particles initially located in a small region can diffuse to a larger
volume\footnote{The opposite can also happen, but our sub-box volume choices
automatically account for such situations.}. In other words, a big region with
enough volume contains almost all information of its central subregion at the
initial time. More details about the data can be found in Section
\ref{sec:data}.

\begin{figure}
\centering
\includegraphics[width=0.45\textwidth]{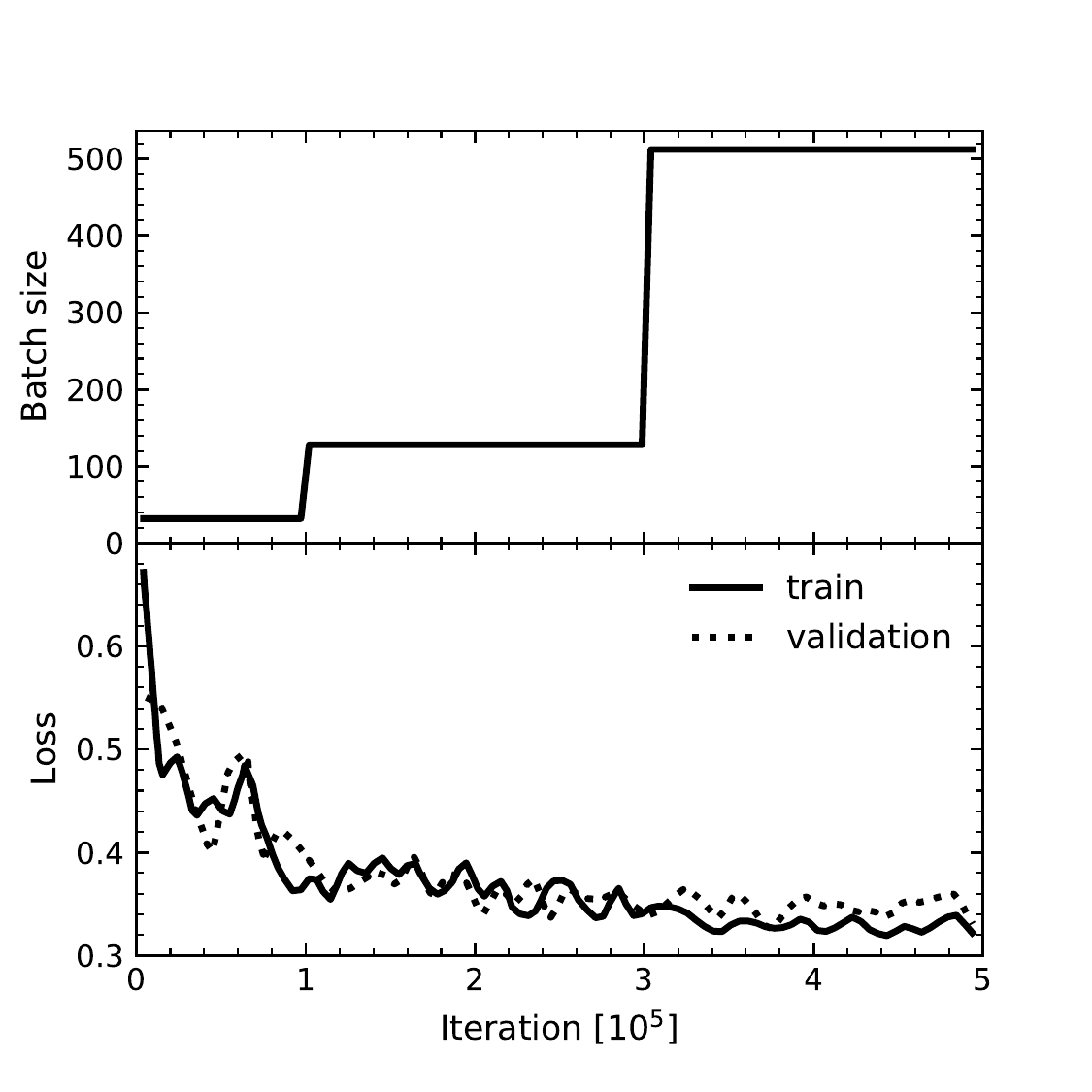}
\caption{\label{fig:training} The batch size (upper panel) and loss functions
    (lower panel). The loss functions are normalized by the variance of initial
    conditions, cf.~Eq.~\eqref{eq:norm_loss_func} and are shown in solid and
    dotted lines for training and validation set, respectively. The loss
    functions in both the training and the validation sets decrease with
    training progress, and there is no obvious over-fitting.
}
\end{figure}

\subsection{Training}
\label{sec:training}

Once the network architecture is fixed, training can help us optimize the
random parameters to suitable values. In the training process, there are some
hyper-parameters which should be selected, such as in the activation function
$a$ and the loss function. The search for the optimal hyper-parameters is
called \textit{fine-tuning}, which requires to train the network multiple times
with different hyper-parameters \citep{Goodfellow2016}. In this paper, we
report our preliminary results of BAO reconstruction by using a neural network,
for which the hyper-parameters are selected roughly and we defer further
fine-tuning to a future study.  Below, we list the hyper-parameters that are
used: 
\begin{enumerate}
\item We initialize the parameters following \cite{variance_scaling}, using the
    function \verb variance_scaling_initializer  in TensorFlow
    \citep{Tensorflow}.  The xavier initialization \citep{xavier} also worked
    well in our test.
\item The learning rate is a hyper-parameter that controls how much we adjust
    the trainable parameters based on the loss gradient.  Reducing the learning
    rate helps to depress the gradient noise, which makes the network tend to
    converge to a local or global minimum. However, the gradient noise can also
    be beneficial in some cases, such as helping to escape ``sharp minima"
    \citep{Smith2017}. In general, the algorithm calculates the gradient in a
    mini-batch, i.e., by computing the gradient against more than one training
    data point, but less than the full dataset, at each iteration. As shown in
    \cite{Smith2017B}, in the training, decreasing the learning rate is usually
    equivalent to increasing the mini-batch size (hereafter, batch size).
    Therefore, we fixed the learning rate to $0.0001$ and changed the batch
    size as shown in Fig.~\ref{fig:training}. 
\item Batch normalization has become a part of the standard toolkits recently
    for accelerating and improving the training of deep network by reducing the
    internal covariate shift \citep{BN}. However, since batch normalization
    uses the mean and variance values in the mini-batch, it is not suitable for
    small or non-i.i.d. mini-batch training \citep{ReBN}. In our task, on the
    one hand, we use sub-box density fields (see Section \ref{sec:data}) when
    training the network which are non-i.i.d. datasets. On the other hand, the
    memory usage of 3-D convolutions limits the batch size of our network.
    Therefore, we remove all batch normalization layers in our network.
\item We use the moment-based Adam Optimizer \citep{Adam} in this work.
\end{enumerate}

In the training, the loss function is the most important indicator. It can be
used to monitor the network's ability and over-fitting. Here, over-fitting
means the network is trained to work so well on the training set that it works
poorly on data it has not seen before. The training set is the dataset fed into
the network and used to calculate the gradients for updating trainable
parameters. In the training, the data that the network ``has not seen before''
is called the validation set. The validation set will also be fed into the
network, but only its loss will be used to indicate over-fitting or not. When
the network is over-fitted, the loss of training set reduces but the validation
loss increases instead.

In Fig.~\ref{fig:training}, we show the batch size and loss function in
training. In the top panel, the batch size increases gradually with the
training progress to reduce the gradient noise. In the bottom panel, the loss
functions of the training set and validation set are represented by black and
red solid lines, respectively. The loss function shown here is normalized by
the variance of initial conditions, in other words
\begin{equation}\label{eq:norm_loss_func}
    \frac{\mathrm{MSE}}{\sigma^2}=
        \frac{\sum^{k}\left(
            f(\delta_{\fc}^k;\boldsymbol{\theta})
            -\delta_{\ic}^k\right)^2}{\sum^{k}\left(\bar{\delta_{\ic}}-\delta_{\ic}^k\right)^2},
\end{equation}
where $\bar{\delta}_{\rm i}$ denotes mean of the initial density field. In this
case, the loss will be $1$ if the network predicts initial density only by its
mean value. We find the losses in both the training set and the validation set
decrease gradually in the training, and there is no obvious over-fitting. Thus,
we do not use regularizations such as $L2$ regularization or dropouts
\citep{dropout} in our network, aimed to avoid over-fitting.  We tested adding
residual architectures \citep{He2015} to the network as well, but found no
significant improvement; this may be because our network is not very deep.
Therefore, we did not use residual layers in our model.

\subsection{Data set}
\label{sec:data}

The dataset in this study is based on the Indra simulations (Falck et al, {\it
in preparation}), a suite of N-body simulations (512 runs) evolved from
different initial conditions using L-Gadget \citep{Gadget2}, each with
$1024^{3}$ dark matter particles in a periodic cube of 1~$h^{-1}\mathrm{Gpc}$
on a side. The cosmological parameters in these simulations are taken to be the
best-fit parameters of WMAP7 \citep{WMAP7}: $\Omega_{m}=0.272$,
$\Omega_{\Lambda}=0.728$, $\Omega_{b}=0.045$, $h=0.704$, $\sigma_{8}=0.81$, and
$n_{s}=0.967$, where $\Omega_m, \Omega_\Lambda$ and $\Omega_b$ are respectively
the present-day density parameters for matter, cosmological constant and
baryons; $h=H_0/(100~{\rm km/s/Mpc})$, $\sigma_8$ is the rms matter density
fluctuations at $z=0$, and $n_s$ the index of the primordial power spectrum of
the density perturbations.

In total, 24 simulations were used to build the dataset, equally split between
training, validation and test sets.  We define the snapshot $z=10$ as the
initial condition and the snapshot $z=0$ as the final condition. Here, we
choose the initial time arbitrarily: if we instead define the initial time as a
redshift higher than $10$, we only need to retrain the network with the data at
{that} corresponding redshift.

For each simulation, we assign the dark matter particles onto a $512^{3}$ grid
using Piecewise Cubic Spline \citep[PCS, see {\it e.g.},][]{PCS}, and smooth
the initial density field with a $3$ $\mpch$ Gaussian filter. The network is
designed to input $39^3$ cells and output one value for reducing computational
complexity. Here, the input is a cubic region of the final density field, and
the output is the predicted density $\delta_{\rm i}$ in the central
$(i,j,k)=(20,20,20)$ cell of this $39^3$ cube.  Since the side length of each
cell is $1.95\mpch$, the input is a cubic sub-box with side length $76\mpch$.

For the training and validation sets, we separate each simulated final density
field into sub-boxes in stride of $16$ grids per side. Thus, in both the
training and validation sets, we generate $32768$ sub-boxes per simulation and
$262144$ sub-boxes in total. When training, we use all sub-boxes in the
training set but randomly select $4096$ sub-boxes from the validation set to
monitor the over-fitting.

To further enlarge the training set, we augment each sub-box with $6$ different
rotations and $8$ different axis-reflections to expand the training set by a
factor of $48$ \citep{Ravanbakhsh2017}. Note that expanding the training set
with more simulations is an alternative possibility. However, the sub-box data
of each simulation occupies a huge storage space (about $14$ GB) in our method,
thus Expanding the data set by using more simulations would have caused storage
pressure; data augmentation is more efficient. Additionally, since over-fitting
is not an urgent problem as shown in Fig.~\ref{fig:training}, more simulations
are not necessary in our study.

Besides the training set and the validation set, an independent test set is
also needed to test the final results of our model, because even though the
validation set makes no contribution to the gradient, it is used to choose the
hyper-parameters. Unlike the training and validation sets, we do not use the
test set to measure the loss.  In the test results, the network is seen as a
complex convolutional kernel and we convolve it on the whole final density
field. In this way, we get the corresponding reconstructed density field.  We
note that all results in Section \ref{sec:result} are calculated from $8$
simulations in a test set with the same cosmology, but different initial
conditions compared to the training and validation sets.

\begin{figure*}
\centering
\includegraphics[width=0.75\textwidth]{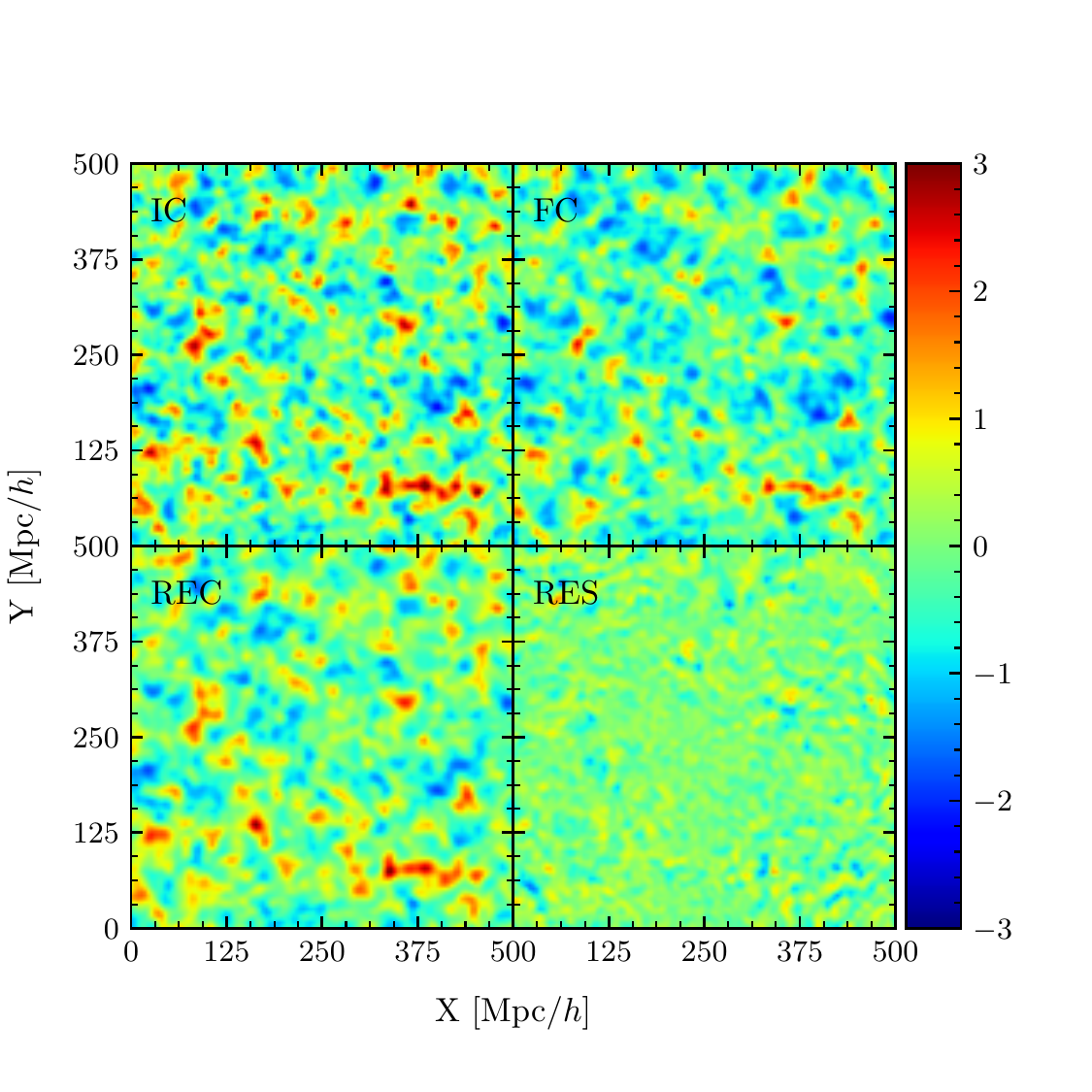}
    \caption{({\it Colour Online}) Density maps. We show the density map in a
        $1.95~\mpch$ slice of IC: initial condition $\delta_{\ic}$, FC: final
        condition $\ln(\delta_{\fc}+1)$, REC: reconstruction $\delta_\rec$ and
        RES: residual between reconstruction and initial condition
        $\delta_{\rec} - \delta_{\ic}$. For clarity, we have linearly
        extrapolated the corresponding density contrast $\delta$ to $z=0$ using
        the linear growth factor, and smoothed all density fields with a
        Gaussian filter with $\sigma=4\mpch$.  }
\label{fig:map}
\end{figure*}

\section{Results}
\label{sec:result}

\subsection{Density maps and probability density functions}

Visualisation of density maps can provide intuition of the quality of the
reconstruction. In Fig.~\ref{fig:map}, we show the density maps of the initial
condition $\delta_{\ic}$, final condition $\delta_{\fc}$, reconstruction
$\delta_{\rec}$, and the residual between reconstruction and the initial
condition, ${\rm RES} \equiv \delta_{\rec} - \delta_{\ic}$, respectively. To
show these density maps clearly, we linearly extrapolate the corresponding
density contrasts $\delta$ to $z=0$ using the linear growth factor $D_+$ and
perform a $4 \mpch$ Gaussian smoothing on all density fields. The projection
depth of all slices is $1.95 \mpch$. As shown in the residual map,
$\delta_{\rec}$ is almost identical to the initial density $\delta_{\ic}$. 

\begin{figure}
\centering
\includegraphics[width=0.45\textwidth]{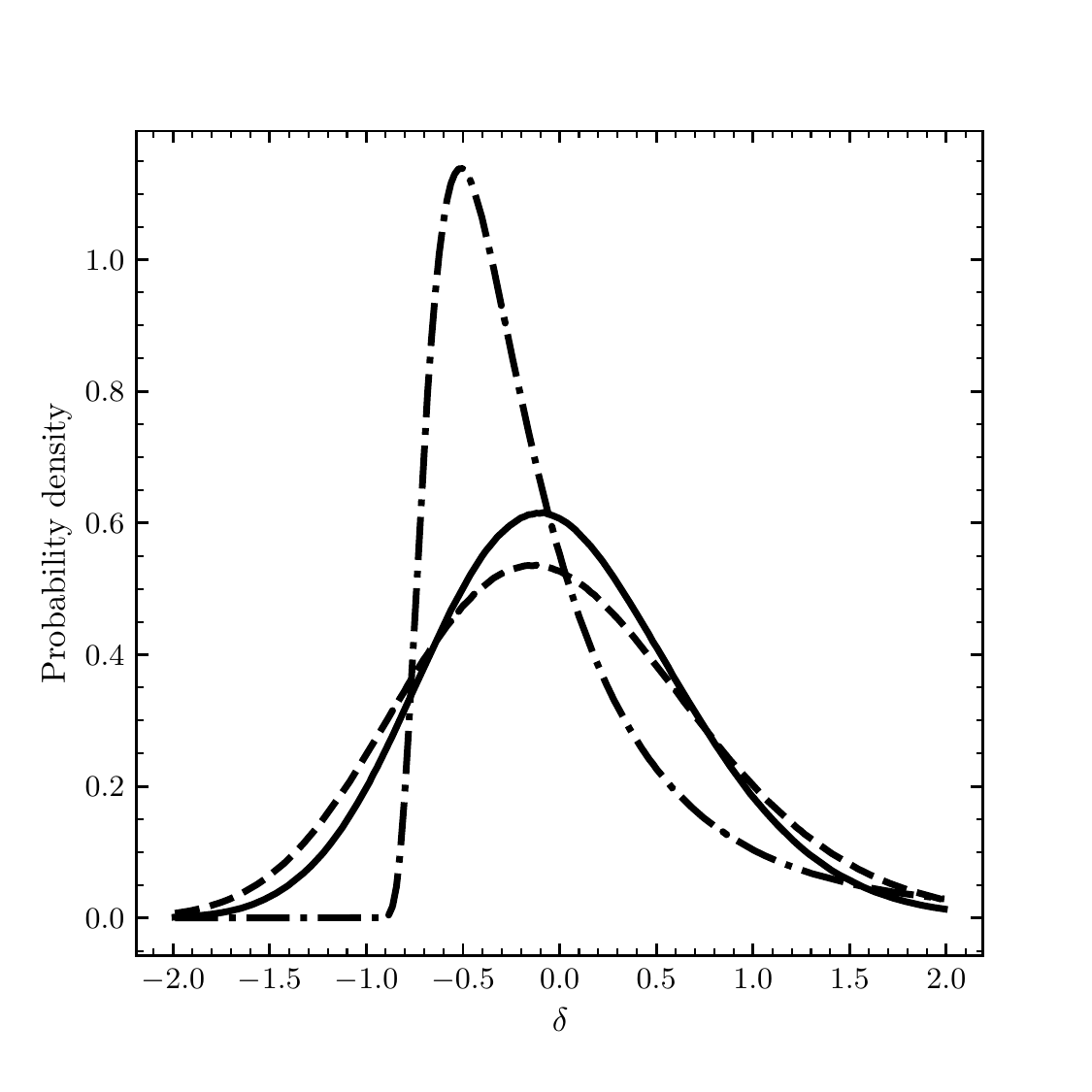}
    \caption{\label{fig:PDF} Probability density functions. The dashed,
        dotted-dashed and  solid lines indicate the PDFs of $\delta_{\ic}$,
        $\delta_{\fc}$ and $\delta_{\rec}$. We have linearly extrapolated the
        initial and reconstructed density fields to $z=0$ with the linear
        growth factor, and perform a $4\mpch$ Gaussian smoothing as in
        Fig.~\ref{fig:map}. Unlike $\delta_{\fc}$, the PDF of $\delta_{\rec}$
    is close to $\delta_{\ic}$.  }
\end{figure}

To further quantify the quality of the reconstruction, we show the probability
density functions (PDFs) in Fig. \ref{fig:PDF}. The dashed, dotted-dashed and
solid lines indicate the PDFs for $\delta_{\ic}$, $\delta_{\fc}$ and
$\delta_{\rec}$, respectively. As in Fig. \ref{fig:map}, we linearly
extrapolate the corresponding density fields to $z=0$ and perform a $4\mpch$
Gaussian smoothing. We find that, compared with the final condition, the PDF of
the reconstruction is much closer to the initial conditions, and have a shape
that is closer to Gaussian.

\subsection{Transfer function}
\begin{figure}

\centering
\includegraphics[width=0.45\textwidth]{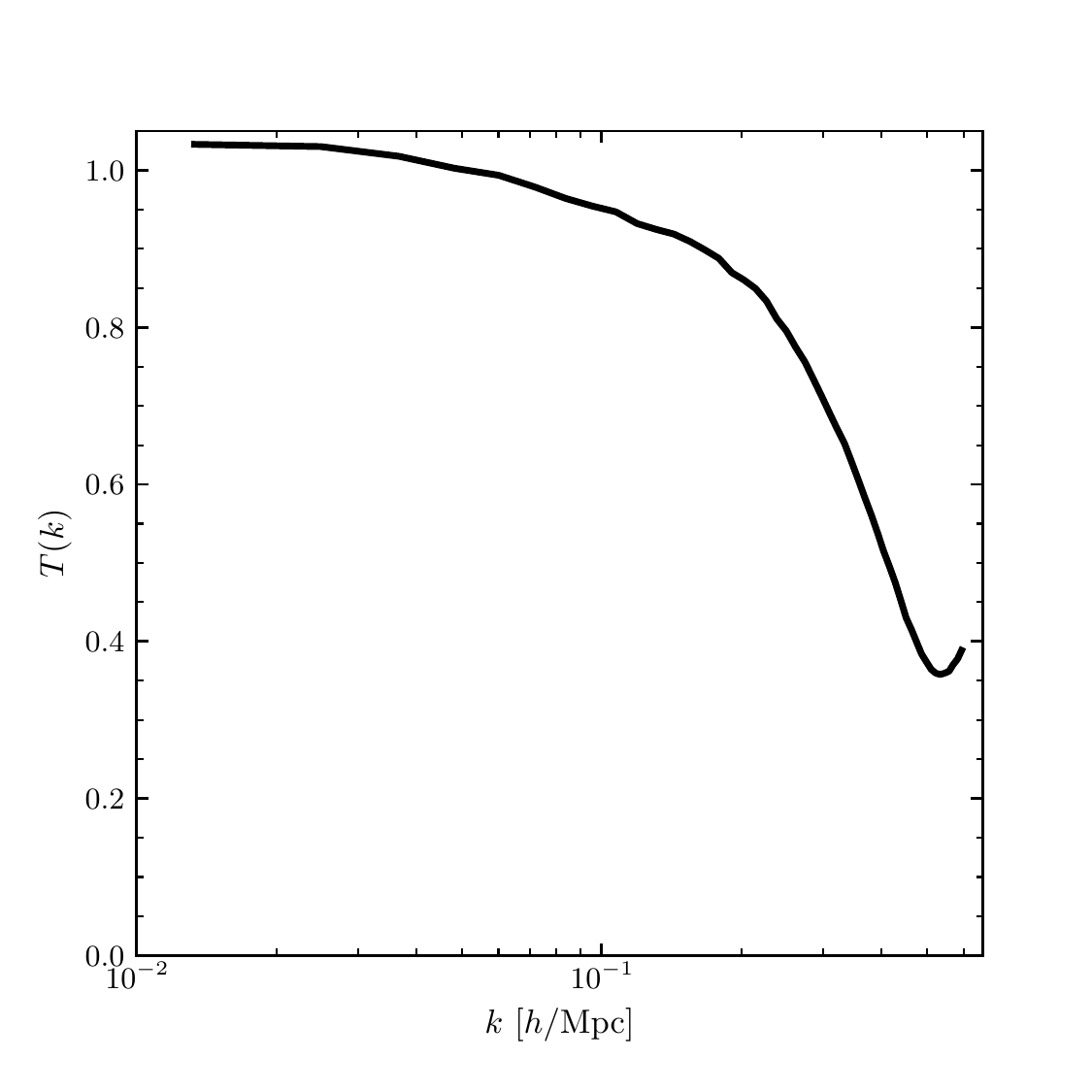}
\caption{\label{fig:TF} The mean transfer function over $8$ simulations. 
For most points, the standard deviations are smaller than 
$5$\textperthousand\ compared to the transfer function, and so they are not
shown in the figure.}
\end{figure}

The transfer function can be used to quantify the discrepancy between the power
spectra of the initial condition and reconstruction. It is defined as
\begin{equation}
\label{eq:TF}
T(k)=\sqrt{P_{\rec}(k)/P_\ic(k)},
\end{equation}
where $P_{\rec}(k)$ and $P_\ic(k)$ are respectively the power spectra of $ \delta_{\rec}$ and $\delta_\ic$. 

Fig.~\ref{fig:TF} shows the transfer function averaged over $8$ simulations.
The transfer function decays towards small scales, which is as expected since
the complicated small-scaling clustering features are harder to reproduce. We
also note that the reconstructed density field is slightly biased on the
largest scales, where the transfer function $T$ is larger than $1$. To test if
this result is due to sample variance, we have measured  the standard
deviations over the 8 simulations, and found that for all scales except the
first $k$ bin shown in Fig.~\ref{fig:TF}, the scatters are smaller than
$5$\textperthousand. Even for the first $k$ bin which suffers most from cosmic
variance,  the standard deviation is only $\sim1.2\%$. Therefore this bias
seems indeed to be systematic.  A possible reason for this is that our method
reconstructs the initial conditions from a small sub-box volume, which lacks
larger-scale information.  Fortunately, these scales on which the bias occurs
are not important in BAO reconstruction, since they are nearly unaffected by
nonlinear evolution and do not need reconstruction.  Furthermore, if we can
measure this bias in simulations and it turns out to stable in different cases,
we can calibrate the density fluctuation by transfer function, as is widely
used in other reconstruction methods
\citep{Schmittfull2017,Zhu2017,Seljak2017}.

\subsection{Correlation coefficient}

\begin{figure}
\centering
\includegraphics[width=0.45\textwidth]{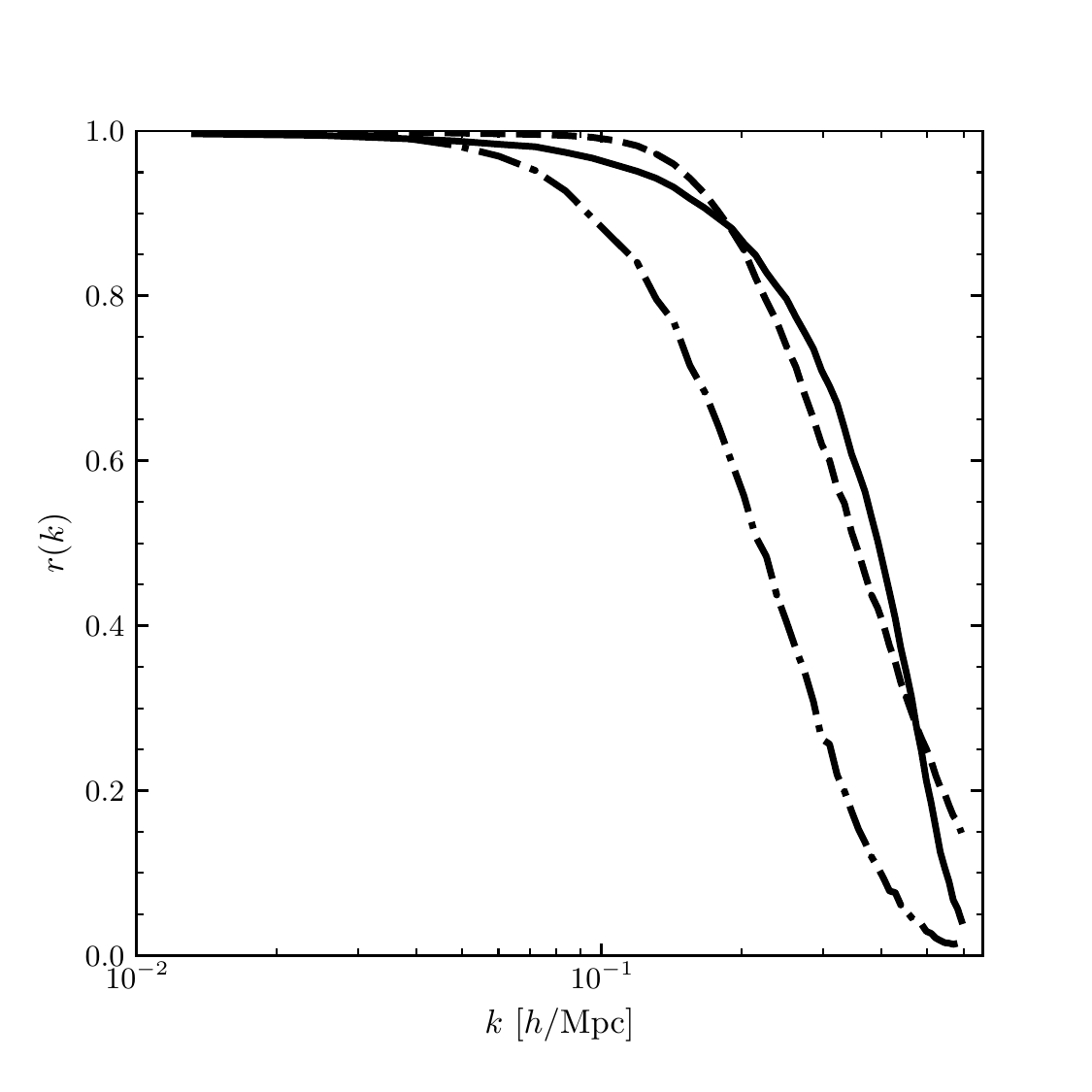}
    \caption{\label{fig:CC} The correlation coefficient between the initial
        condition and the reconstruction (the black solid line). The
        reconstructed density field is about $90\%$ correlated with the initial
        density at $k\simeq0.2\hmpc$ and about $50\%$  correlated at
        $k\simeq0.4\hmpc$. As a comparison, the correlation between the initial
        condition and standard reconstruction is shown as the dashed line.
        Compared with the standard reconstruction, our result is more
        correlated with the ground truth between $k\simeq0.2\hmpc$ and
        $0.5\hmpc$. On scales larger than $0.2\hmpc$, the standard
        reconstruction works better, possibly because of the small sub-box size
        in our method which means that the large-scale information is not used.
    The dotted-dashed line shows the correlation between the initial and final
conditions.}
\end{figure}

The correlation coefficient between two fields describes their correlation in
Fourier phases. It is defined as 
\begin{equation}
\label{eq:CC}
r(k)=P_{12}(k)/\sqrt{P_{1}(k)P_2(k)},
\end{equation}
where $P_1$ and $P_2$ are the auto power spectra of the two fields, and
$P_{12}$ their cross power spectrum. In Fig.~\ref{fig:CC}, we show the
correlation coefficient between the initial condition and reconstruction with
the black solid curve, and use the dotted-dashed curve to denote the
correlation between the initial and final conditions. We can see that the
reconstruction increases $r$ to $\sim0.5$ at $k=0.4\hmpc$. The restoration of
the information larger than this scale is enough to recover the BAO signal,
because the BAO peaks at $k\gtrsim0.4\hmpc$ are weaker than $1$ percent (see
Fig.~\ref{fig:BAO}), not currently detectable in observations.

As comparison, we also show the correlation between the initial conditions and
the standard reconstruction with the dashed curve. Here, the standard
reconstruction is performed using ``Nbodykit'' \citep{nbodykit}, with Gaussian
smoothing on a scale of $20\mpch$. On scales larger than $k=0.2 \hmpc$, the
standard reconstruction method works slightly better, which again could be
because the limited sub-box size in our method removes the clustering
information on those large scales. On scales between $k=0.2\hmpc$ and
$0.4\hmpc$, where the BAO signal is still strong, our method is better
correlated with the initial conditions.

\subsection{BAO signal}

\begin{figure}
\centering
\includegraphics[width=0.45\textwidth]{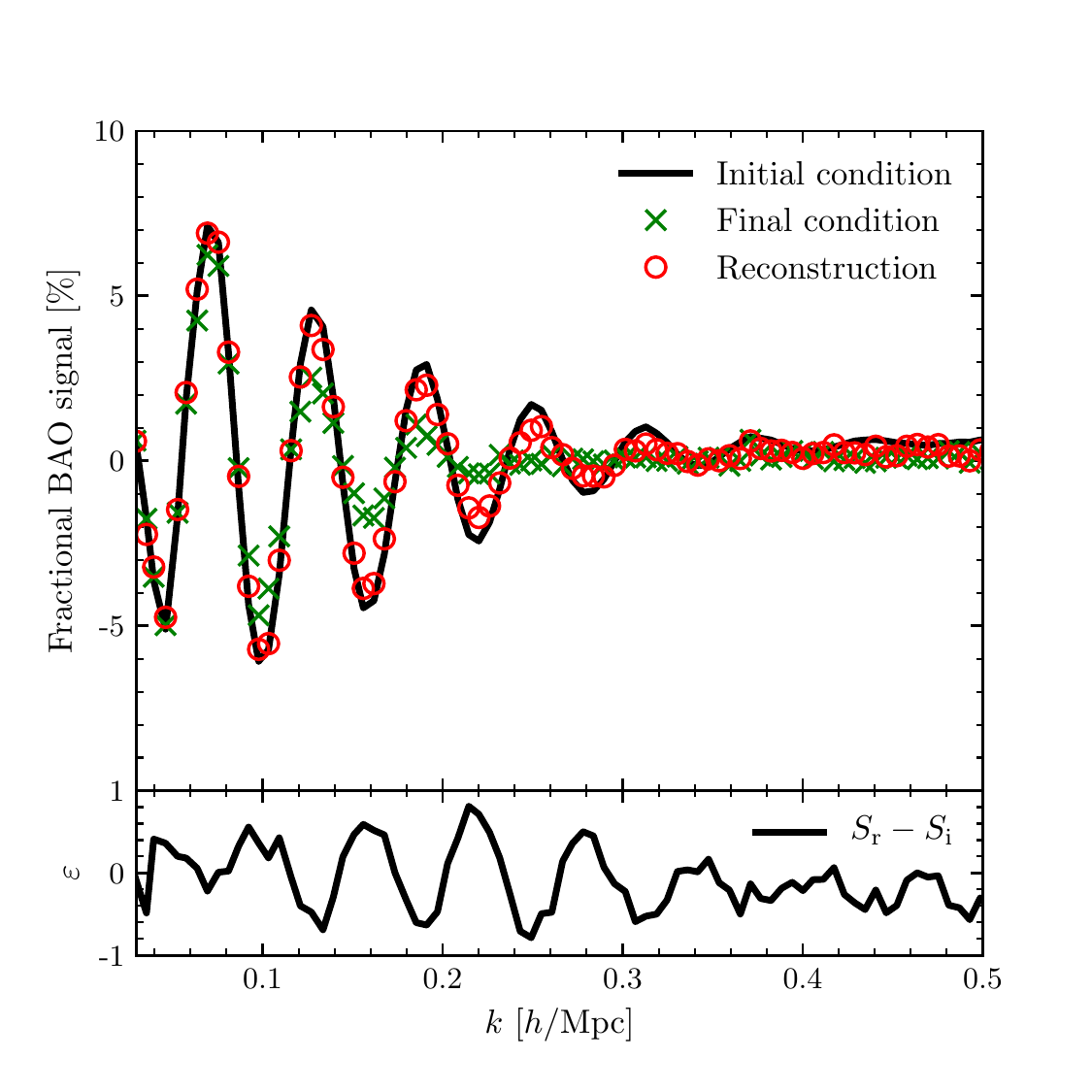}
    \caption{\label{fig:BAO} ({\it Colour Online}) Fractional BAO signals,
        Eq.~\eqref{eq:BAO}, in the power spectrum. To reduce the cosmic
        variance, we run a pair of simulations from two initial conditions with
        the same initial random seed but generated by initial power spectra
        with and without BAO wiggles.  {\it Upper panel}: the black solid line,
        green cross points and red hollow circles indicate, respectively, the
        fractional BAO signals in the initial conditions, final conditions and
        reconstruction. Compared with the final nonlinear case, reconstruction
        increases the signal-to-noise ratio until $k\simeq0.4 \hmpc$. {\it
    Lower panel}: the difference of fractional BAO signals measured from the
initial conditions ($S_{\ic}$) and the reconstruction ($S_{\rec}$).  }
\end{figure}

To directly test the quality of our reconstruction of the BAO signal, we show
the fractional BAO signal in Fig.~\ref{fig:BAO}. The latter is defined as
\begin{equation}
\label{eq:BAO}
S = (P_{\mathrm{wiggle}}-P_{\mathrm{nowiggle}})/P_{\mathrm{nowiggle}}, 
\end{equation}
where the subscripts ``wiggle'' and ``nowiggle'' denote simulations evolved
from initial power spectra with and without BAO wiggles \citep{Vlah2015}. These
simulation pairs have the same initial random seed, which helps to cancel the
cosmic variance in the fractional BAO signal \citep{Schmittfull2017}. In the
top panel, the black solid curve is the fractional BAO signal calculated from
the initial conditions, in which we can clearly see a series of BAO peaks.
However, for the final conditions, shown by the green cross points, the BAO
peaks are broadened, which means that the signal-to-noise ratio of the peaks
decreases because of nonlinear evolution, especially on scales
$0.2$-$0.4\hmpc$. After reconstruction, shown by the red hollow circles, the
signal-to-noise ratio of BAO peaks is improved, until about $k=0.4\hmpc$. We
also show the difference between initial conditions and our reconstruction in
the bottom panel. On all scales, we find the differences of fractional BAO
signals measured from the initial condition and the reconstruction are smaller
than 1 percent. This indicates our reconstruction succeeds in removing the
effect of nonlinear evolution, and recovering the BAO wiggles in the initial
conditions. As Fig.~\ref{fig:BAO} shows, we can recover the peak around $k=0.32
\hmpc$ and partly the peak around $k=0.38 \hmpc$.

\section{Discussion}
\label{sec:discussion}

\subsection{Cosmology dependence}
\label{ssec:WMAP}

In this study, we train our network using simulations with specific
cosmological parameters, which introduces a cosmological dependence into our
model. In this subsection, we check this dependence in more detail. 

To check the cosmology dependence, we run another two pairs of simulations with
a $500\mpch$ box size from initial conditions with and without BAO wiggles like
in Fig.~\ref{fig:BAO}. These simulations use different cosmological parameters
from the training set: in the training set, the Indra simulation use the
best-fitting cosmological parameters of WMAP+BAO+$\rm{H}_0$ based on the 7-year
WMAP results, while in these new simulations, we choose the best-fitting
parameters based on the WMAP-only data in WMAP5 \citep{WMAP5} and WMAP9
\citep{WMAP9}, to increase the difference from the training sample. For
clarity, in the discussion below about cosmology dependence, we call the
universe of the training set the ``training cosmology'' and use ``truth
cosmology'' to indicate both new simulations of WMAP5 and WMAP9 in the
corresponding comparisons.  In the following comparisons, we train our model in
the training cosmology, but use it to reconstruct the BAO signal in the two
truth cosmologies.

\begin{figure}
\centering
\includegraphics[width=0.45\textwidth]{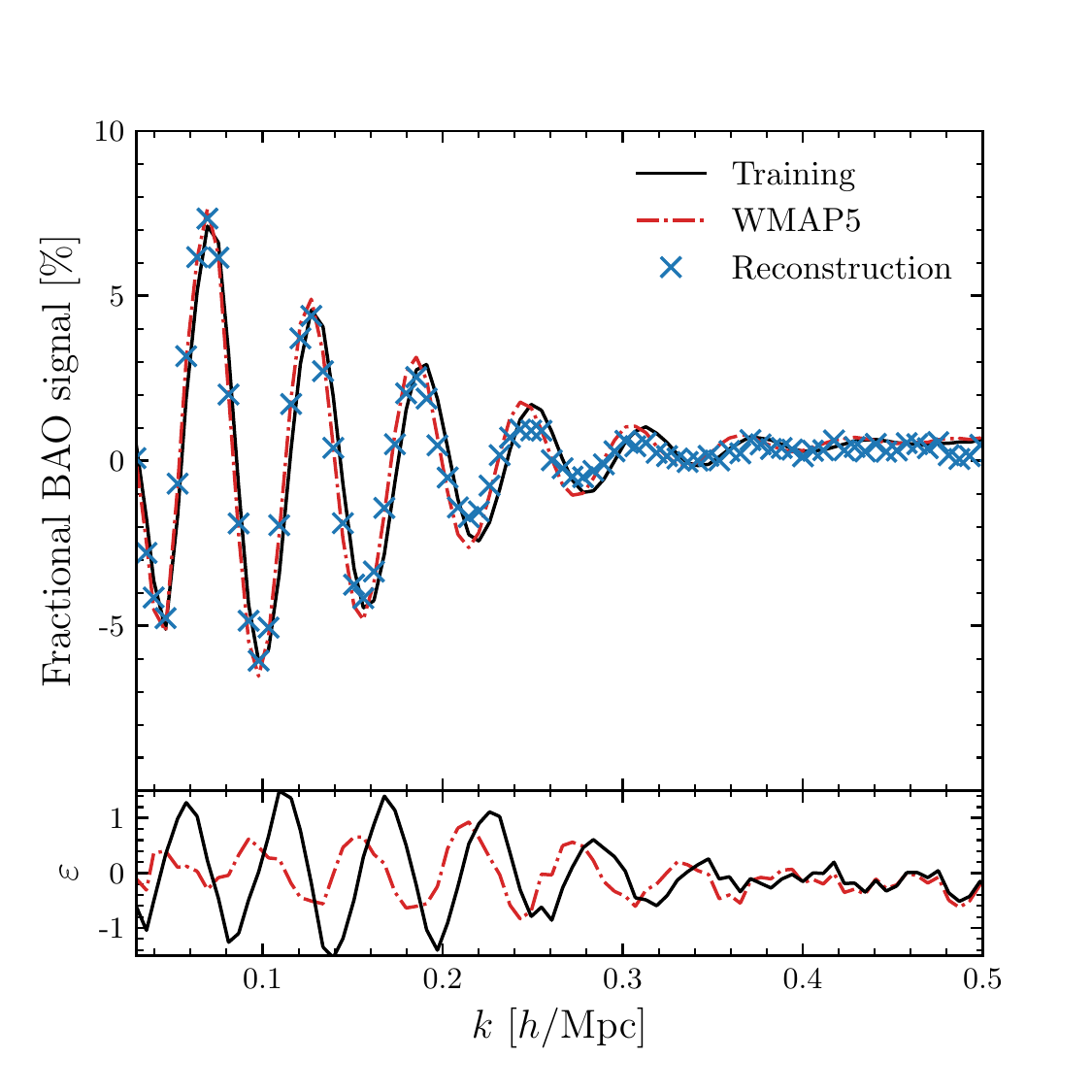}
\caption{\label{fig:WMAP5} ({\it Colour Online}) 
    Cosmology dependence.  The network is trained in the training (WMAP7)
    cosmology, while the reconstruction is applied in the truth (WMAP5)
    cosmology. In the upper panel, the black solid and red dot-dashed lines
    indicate the fractional BAO signals of the initial conditions in the
    training and truth cosmologies, respectively. The cyan crosses show the
    reconstruction in the truth cosmology.  Although the network is trained by
    simulations in the training cosmology, we find the reconstructed signal is
    closer to the truth cosmology than the signal in the training cosmology. In
    the lower panel, the black solid and red dotted-dashed lines show the
    differences between the reconstruction and initial conditions in training
    and truth cosmologies, respectively.  }
\end{figure}

The results from the WMAP5 cosmology are shown in Fig.~\ref{fig:WMAP5}. In the
upper panel, the black solid and red dotted-dashed lines indicate the
fractional BAO signals of the initial conditions in the training and truth
cosmologies, respectively. The reconstruction in the truth cosmology is shown
by the blue cross points. We find our reconstruction is closer to the truth
cosmology than to the training cosmology on almost all scales.  This indicates
that, instead of `remembering' the BAO signal of {\it the} training set,  the
network model has indeed successfully learned the relation
$\delta_{\ic}=f(\delta_{\fc};\boldsymbol{\theta})$, which is model-independent
and allows it to reconstruct the initial condition for general cosmologies.  In
the lower panel, the black solid and red dotted-dashed lines show the
differences between the reconstruction and the initial condition in the
training and truth cosmologies, respectively. It is interesting to note that
the differences are actually smaller for the truth cosmology than for the
cosmology upon which the network has been trained.
 
\begin{figure}
\centering
\includegraphics[width=0.45\textwidth]{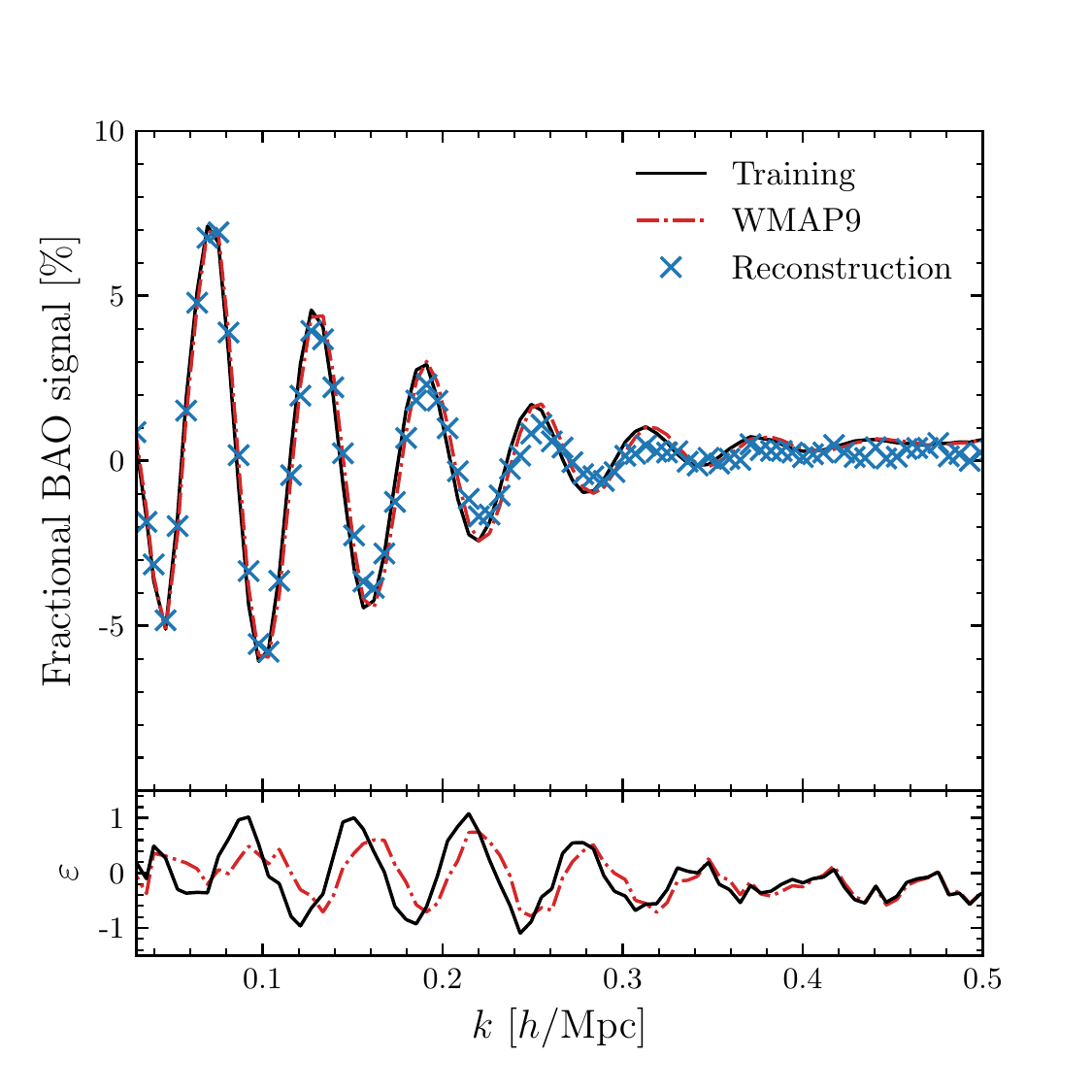}
\caption{\label{fig:WMAP9} ({\it Colour Online}) 
    The same as Fig.~\ref{fig:WMAP5}, but with the WMAP9 cosmology as the truth
cosmology. Since the training and truth cosmologies are so close, there are
only very small differences between the BAO signals of the training, truth and
reconstruction at $k\lesssim0.2\hmpc$, although the reconstruction does agree
better with the truth cosmology as expected.}
\end{figure}

Fig.~\ref{fig:WMAP9} is similar to Fig.~\ref{fig:WMAP5}, but shows a second
test of cosmology dependency using WMAP9 as the truth cosmology. Since the
cosmological parameters of the training and truth cosmologies are very close in
this case, there is only a slight discrepancy of the BAO signals in those two
cosmologies. We again find that the reconstruction is closer to the truth
cosmology, although there are only very small discrepancy between the two,
especially on scales between $0.1$ and $0.3 \hmpc$.

\begin{figure}
\centering
\includegraphics[width=0.45\textwidth]{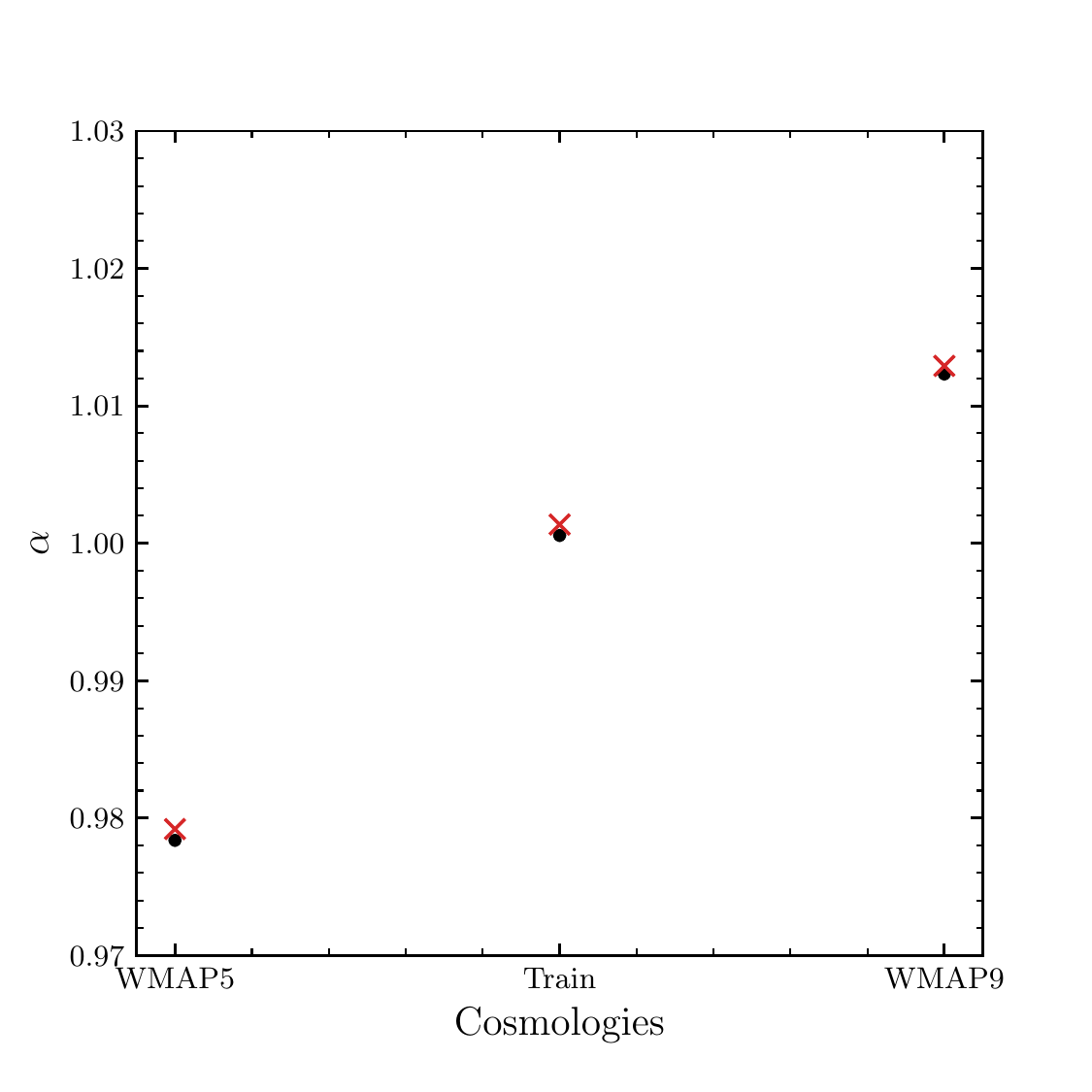}
    \caption{\label{fig:cosms} ({\it Colour Online}) The scale dilation
        parameter $\alpha$, cf.~Eq.~\eqref{eq:scale_dilation}, which quantifies
        the shift of the BAO peak position in a given density field with
        respect to the peak positions in the initial condition of the training
        cosmology.  The $x$-axis shows the three different cosmologies used in
        our tests, and the black points and red crosses indicate the best-fit
        $\alpha$ values from the initial condition and reconstruction,
        respectively. The difference between the BAO peak positions in the
        initial condition and the reconstruction for both truth cosmologies
        (WMAP5 and WMAP9) is much smaller than the difference between its
        locations in the different cosmologies. This is consistent with the
        results shown in Figs.~\ref{fig:WMAP5} and \ref{fig:WMAP9}.
}
\end{figure}

To quantify whether this method can be used to distinguish between different
cosmologies, we define a scale dilation parameter $\alpha$ that is used to
adjust the location of the BAO peaks, as 
\begin{equation}\label{eq:scale_dilation}
 S(k) = S_{\mathrm{t}}(k/\alpha),
\end{equation}
where the $S$ is the fractional BAO signal defined in Eq.~(\ref{eq:BAO}) and
$S_{\mathrm{t}}$ indicates the initial fractional BAO signal in the training
cosmology. Since the definition of the fractional BAO signal has removed most
of the cosmic variance and the reconstruction has removed nonlinear damping,
the parameter $\alpha$ will show how much the peak location of $S$ is shifted
with respect to $S_{\mathrm{t}}$.  We fit $\alpha$ with $S$ being the
fractional BAO signal of the reconstruction and initial condition respectively,
for the training, WMAP5 and WMAP9 cosmologies.  In Fig.~\ref{fig:cosms}, the
black points indicate the $\alpha$ fitted from the initial conditions and the
red cross points are fitted from the reconstruction.  Note that $\alpha=1$ by
definition for the initial condition of the training cosmology (the second
black dot). There is a (very) slight shift of the reconstructed BAO peaks even
in the training cosmology ($\alpha>1$ for the second red cross), and this shift
seems to be the same for the reconstruction results of the two truth
cosmologies, indicating again that the trained network has negligible cosmology
dependence\footnote{The shift itself could be a sign that reconstruction cannot
fully remove the broadening of the BAO peaks caused by nonlinear structure
formation.}. More importantly, this shift is much smaller than the difference
in $\alpha$ that is due to the underlying cosmology (which is at the percent
level), suggesting that different models can be distinguished between our
reconstruction method.

It should be noted that, to overcome the cosmology dependency completely, we
should train the network with simulations in a series of cosmological
parameters as done by \cite{Ravanbakhsh2017}. A more detailed analysis of this
is beyond the scope of the present paper and will be left for future work.

\subsection{Boundary effects}
\label{ssec:boundary}

\begin{figure*}
\centering
\includegraphics[width=1.0\textwidth]{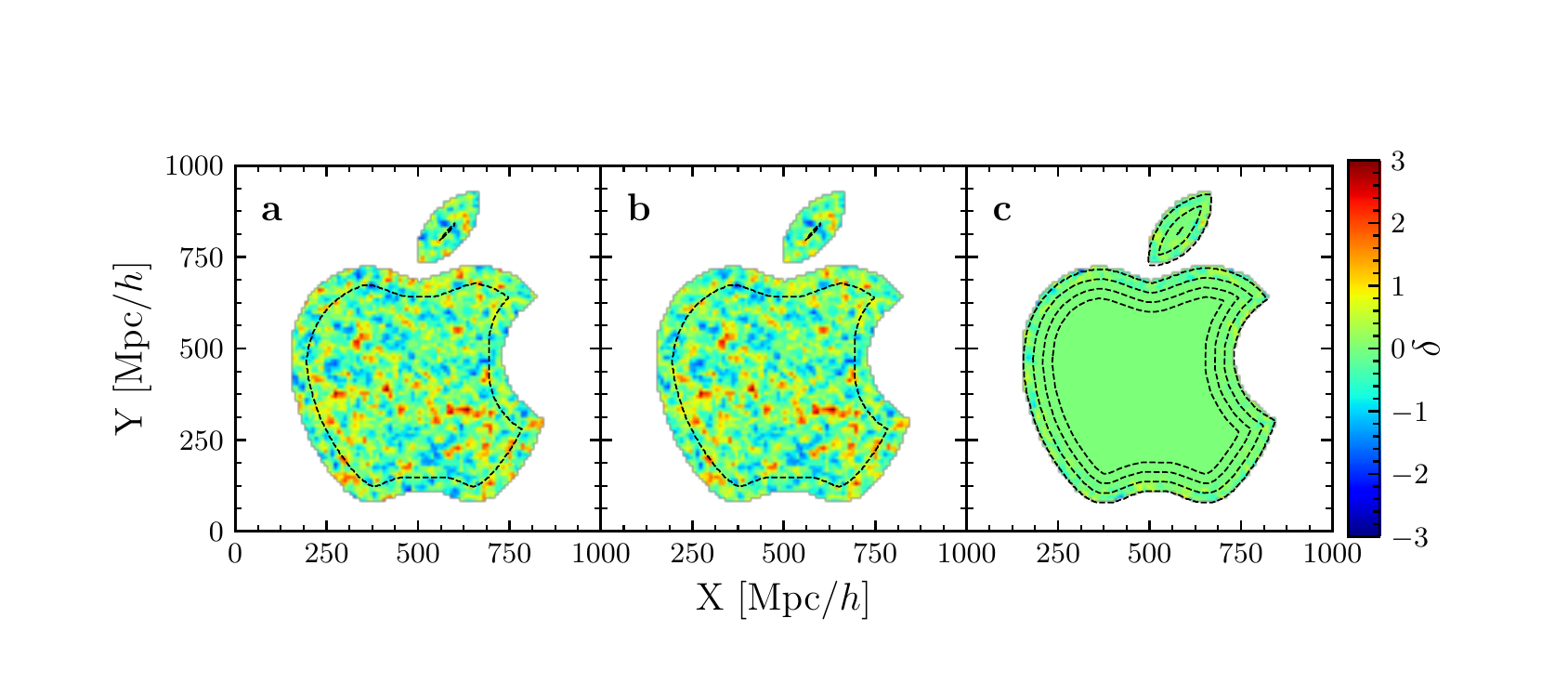}
    \caption{\label{fig:mapb} ({\it Colour Online})
    (a) The density map of the full reconstruction. (b) The density map of the
    boundary reconstruction. (c) The residual between the full and boundary
    reconstructions. In all panels, only the areas located in the assumed
    survey boundary are shown. In Panels (a) and (b), the black dashed lines
    indicate the \Crit, while in Panel (c), from outside to inside, the dashed
    lines indicate the positions at which $d_{\mathrm{b}}$ is $0$, $26.2$,
    $52.3$ and $78.5 \mpch$, respectively. With $d_{\mathrm{b}}$ decreasing,
    the residual becomes increasingly stronger due to the missed information
    outside the boundary.
}
\end{figure*}

The discussion so far has been in the idealized case of dark matter, with no
boundary effects. BAO measurement in galaxy survey data faces other
complications, such as redshift-space distortions, shot noise, galaxy bias, and
survey boundaries. Most perturbation based reconstruction algorithms estimate
the displacement field in Fourier space. In this case, the irregular survey
boundary and incomplete information near it disturbs the reconstruction results
as far as $100\mpch$ from the boundary \citep{Zhu2019}.

In order to deal with the irregular survey boundary, one can build a forward
model that includes the boundary, and optimize them together \citep[see,
\eg][]{Seljak2017,Feng2018,Modi2018}.  If we only focus on the effects of
survey boundary, a forward-modeling method is a suitable choice because the
boundary is fixed and we know it very well. However, limited by computing power
and memory, the forward model usually cannot be too complicated. In practice, a
limited forward model may be hard to generalize, and there is a trade off
between variance and bias, \eg much effort is required to understand the bias
of the forward model and make it close to the actual problem. The survey
boundary is easy to handle in forward modeling, though not all observational
effects are.

Since our network model estimates the initial conditions from a sub-box instead
of the global density field, we expect that survey boundaries should have a
small impact in this method.  In this subsection, we check such a boundary
dependence of our reconstruction (hereafter, boundary reconstruction) by
assuming a survey boundary like the Apple logo, which provides a couple of
separated irregular regions.  For comparison, we use ``full reconstruction'' to
indicate the reconstruction that use all information in the box (i.e., with
periodic boundaries).

Fig. \ref{fig:mapb} shows the density maps of (a) the full reconstruction,  (b)
the boundary reconstruction, and (c) the residual between them. All these maps
are shown in a $1.95 \mpch$ slice. For the boundary reconstruction, we fill all
cells outside the assumed survey boundary with the cosmic mean density during
reconstruction. For clarity, we define a boundary distance $d_\mathrm{b}$ which
describes the nearest distance to the survey boundary of each cell, which will
be helpful for quantifying the impact of survey boundary in reconstruction. In
our method, the area not affected by the survey boundary (hereafter, \Crit) is
circled by the black dashed line in the full and boundary reconstruction maps.
The $d_{\mathrm{b}}$ of \Crit\ ranges from $37 \mpch$ to $52.3 \mpch$,
depending on the position in the map, because of the sub-box length used in
reconstruction. In the residual panel, from outside to inside, the dashed lines
indicate $d_{\mathrm{b}}$ as $0$, $26.2$, $52.3$ and $78.5 \mpch$,
respectively. Only a slight discrepancy between full and boundary
reconstruction maps is found in the area $0<d_{\mathrm{b}}<26.2 \mpch$.

\begin{figure}
\centering
\includegraphics[width=0.45\textwidth]{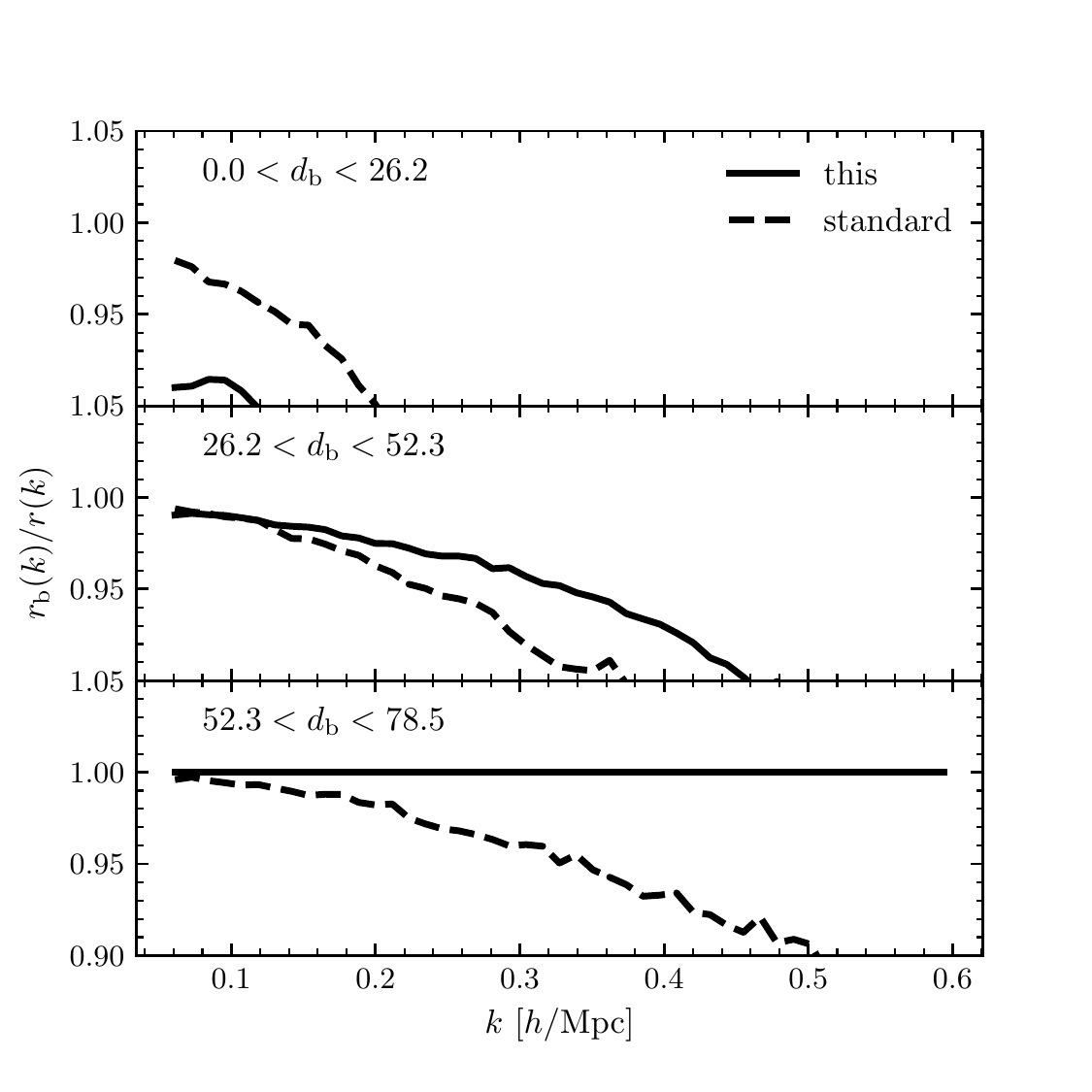}
    \caption{\label{fig:CCb} The ratio between the correlation coefficients
        from the boundary and full reconstructions. The solid and dashed lines
        indicate our method and the standard reconstruction method,
        respectively. From top to bottom, different ranges of the boundary
        distance $d_{\rm b}$ are used to define the region that is used to
        calculate the correlation coefficient, as indicated in the legends.
        These correspond to the three regions between neighboring dashed lines
        in the right panel of Fig.~\ref{fig:mapb}.  }
\end{figure}

In Fig.~\ref{fig:CCb}, we show the ratio of correlation coefficients between
boundary and full reconstruction in areas of $0<d_{\mathrm{b}}<26.2 \mpch$,
$26.2<d_{\mathrm{b}}<52.3 \mpch$ and $52.3<d_{\mathrm{b}}<78.5\mpch$,
respectively. Note that to calculate the correlation coefficients in these
three areas, we mask out the rest of the field.  The solid curves indicate the
results of our method, and the dashed curves are for the standard
reconstruction method. To implement boundary reconstruction in the standard
method, we replace the matter distribution beyond the survey boundary with a
random catalog. As $d_{\mathrm{b}}$ increases, in both methods, the correlation
coefficients of the boundary reconstruction are less and less affected by the
survey boundary. In the upper panel, where the boundary distance is smaller
than $26.2 \mpch$, the correlation coefficient ratios in both methods decline
quickly, reaching 90\% at $k\simeq0.2 \hmpc$. In the middle panel, where
$d_{\mathrm{b}}$ is close to the \Crit, the difference between boundary and
full reconstructions in our method is significantly reduced down to
$k\simeq0.46 \hmpc$. When we calculate the correlation coefficient in region of
$52.3<d_{\mathrm{b}}<78.5 \mpch$, the ratio between boundary and full
reconstruction of our method is equal to one because this region is located
entirely inside the \Crit\ (lower panel).

While boundary reconstruction based on the standard reconstruction method works
quite well, especially for $d_{\rm b}<26.2\mpch$, our method still offers
better consistency between boundary and full reconstructions in the region of
$26.2 < d_{\mathrm{b}} < 78.5 \mpch$.  Thus, at a small price of a degraded
performance (compared with standard reconstruction) in the outermost layer near
the survey boundary --- where both methods perform rather poorly anyway --- our
method leads to improved reconstruction results further away from the survey
edge, where boundary effects are present in the standard method.

\section{Conclusion}
\label{sec:con}

We present a new method of BAO reconstruction based on deep convolutional
neural networks, and report its first results when applied to simulated dark
matter density fields.  The objective of reconstruction is to undo the bulk
motions of matter, which could dampen and broaden the BAO peaks that are
present in the primordial matter density field. Therefore, an indicator of its
performance is the phase correlation coefficient between the initial and
reconstructed density fields. Our method can lead to a correlation coefficient
of about $90 \%$ at $k=0.2\hmpc$ and $50\%$ at $k=0.4\hmpc$.  For the
fractional BAO signal, the reconstruction can improve the signal-to-noise ratio
down to $k=0.4\hmpc$, extending the range of scales at which the power spectrum
matches linear theory by a factor about $2$ compared to final condition. 

Since the network is trained by simulations with specific cosmological
parameters, we have checked the cosmology dependency by applying the trained
model to two different cosmologies. We do not find evidence for cosmology
dependence, and the method seems insensitive to the training cosmology. We also
demonstrate that it can distinguish the different cosmologies considered.
However, we caution that the cosmologies on which the method is tested are both
rather close to the training cosmology.  In the future, we will further this
analysis by using a wider parameter range in the training and test sets.

Because its input data is the nonlinear density field in cubic sub-boxes (which
we have chosen to have a side length of $76\mpch$), this new method is by
design robust against boundary effects since areas inside the \Crit{} ($37\mpch
\sim 52.3\mpch$ from the survey edge) are not affected at all. This is an
advantage over the standard reconstruction method, because survey boundaries
can substantially impact BAO reconstruction in galaxy surveys.  Our tests show
that, compared with the standard reconstruction method, the new method improves
the consistency between boundary and full reconstructions in the region of
$26.2 < d_{\mathrm{b}} < 78.5 \mpch$ from the survey edge.

In this paper, we have tested our new scheme in dark-matter-only simulations,
and found that it can accurately remove nonlinear effects on scales larger than
$k=0.4 \hmpc$, and enable us to recover the BAO wiggles {up to} $k=0.32\hmpc$.
In the future, we will test this method by applying it to density field
reconstruct using galaxy surveys \citep{Wang2009}. We expect that using all
{available} galaxies in a survey {can} put {optimally tight} BAO constraints on
the underlying cosmological model.

\section*{Acknowledgements}
We thank Houjun Mo and Huiyuan Wang for very useful discussions. We also thank
the anonymous reviewer for his/her many suggestions for improving this paper.
J.W. acknowledges the support from the National Natural Science Foundation of
China (NSFC) grant 11873051. This work is supported by Astronomical Big Data
Joint Research Center, co-founded by National Astronomical Observatories,
Chinese Academy of Sciences and Alibaba Cloud.

\section*{Data availability}
The data underlying this article will be shared on reasonable request to the corresponding author.



\bibliographystyle{mnras}
\bibliography{refs}




\bsp	
\label{lastpage}
\end{document}